\documentclass[sn-mathphys-num]{sn-jnl}

\usepackage{graphicx}%
\usepackage{multirow}%
\usepackage{amsmath,amssymb,amsfonts}%
\usepackage{amsthm}%
\usepackage{mathrsfs}%
\usepackage[title]{appendix}%

\theoremstyle{thmstyleone}%
\theoremstyle{thmstyletwo}%

\theoremstyle{thmstylethree}%

\begin{document}

\title[Article Title]{The Impact of Knowledge Silos on Responsible AI Practices in Journalism}


\author*[1, 7]{\fnm{Tomás} \sur{Dodds}}\email{t.dodds.rojas@fsw.leidenuniv.nl}
\author[1]{\fnm{Astrid} \sur{Vandendaele}}\email{a.vandendaele@hum.leidenuniv.nl}
\author[2]{\fnm{Felix M.} \sur{Simon}}\email{felix.simon@politics.ox.ac.uk}
\author[3]{\fnm{Natali} \sur{Helberger}}\email{N.Helberger@uva.nl }
\author[4]{\fnm{Valeria} \sur{Resendez}}\email{v.d.c.resendezgomez@uva.nl}
\author[5, 6]{\fnm{Wang Ngai} \sur{Yeung}}\email{justin.yeung@oii.ox.ac.uk}

\affil*[1]{\orgdiv{Leiden University Centre for Linguistics}, \orgname{Leiden University}}

\affil[2]{\orgdiv{Reuters Institute for the Study of Journalism}, \orgname{University of Oxford}}

\affil[3]{\orgdiv{Institute for Information Law}, \orgname{University of Amsterdam}}

\affil[4]{\orgdiv{Amsterdam School of Communication Research}, \orgname{University of Amsterdam}}

\affil[5]{\orgdiv{Oxford Internet Institute}, \orgname{University of Oxford}}

\affil[6]{\orgdiv{Network Science Institute}, \orgname{Northeastern University London}}

\affil[7]{\orgdiv{Berkman Klein Center For Internet \& Society}, \orgname{Harvard University}}

\abstract{The effective adoption of responsible AI practices in journalism requires a concerted effort to bridge different perspectives, including technological, editorial, journalistic, and managerial. Among the many challenges that could impact information sharing around responsible AI inside news organizations are knowledge silos, where information is isolated within one part of the organization and not easily shared with others. This study aims to explore if, and if so, how, knowledge silos affect the adoption of responsible AI practices in journalism through a cross-case study of four major Dutch media outlets. We examine the individual and organizational barriers to AI knowledge sharing and the extent to which knowledge silos could impede the operationalization of responsible AI initiatives inside newsrooms. To address this question, we conducted 14 semi-structured interviews with editors, managers, and journalists at \textit{de Telegraaf}, \textit{de Volkskrant}, the \textit{Nederlandse Omroep Stichting} (NOS), and \textit{RTL Nederland}. The interviews aimed to uncover insights into the existence of knowledge silos, their effects on responsible AI practice adoption, and the organizational practices influencing these dynamics. Our results emphasize the importance of creating better structures for sharing information on AI across all layers of news organizations.}

\keywords{Knowledge silos, responsible AI, journalism, artificial intelligence, information sharing}

\maketitle
\newpage


\section{Introduction}
As the social impact of artificial intelligence (AI) is growing, so do concerns around safety, fairness, responsibility, and the ethics related to the use of these technologies \cite{Helberger2022Towards} \cite{Moran2022Robots}. As \cite{Lu2023Responsible} argue, “compared to traditional software systems, AI systems involve a higher degree of uncertainty and more ethical risk due to their dynamic, autonomous and opaque decision-making and historical-data-dependent behaviors” (p. 1). As AI systems become more integrated into the workflows of newsrooms, the mechanics of editorial decision-making and content curation also risk becoming less transparent \cite{Carabantes2023Why}. To address these challenges, government, academia, industry, and civil society stakeholders are developing best practices and governance frameworks to ensure AI’s responsible development and deployment \cite{Becker2023Policies}\cite{Helberger2022}

While these conversations are taking place, recent studies have shown that news workers are struggling with both technical and cultural challenges around the integration of AI in newsrooms \cite{GutierrezLopez2023Question}. On the one hand, as \cite{Simon2022Uneasy} argues, financial limitations and technical difficulties make it challenging for news organizations to develop AI-based technologies without relying on US-based platform companies and other external vendors who control much of the current AI infrastructure. On the other hand, \cite{Beckett2023Generating} show how media managers are not only struggling with technical issues that stem from a lack of resources. Instead, as they put it, “mitigating AI integration challenges […] requires bridging knowledge gaps that exist among various teams in the newsrooms, a challenge that is more consistent across the board” (p. 33). This lack of information sharing, knowledge gaps, and different internal priorities can not only make it difficult to integrate AI in the first place – but it could also become an obstacle to the responsible implementation of AI systems.

It is in this context that the expanding field of responsible AI in journalism emphasizes the necessity of ethical and accountable AI systems development, deployment, and usage inside newsrooms \cite{Trattner2022Responsible}. Rather than exclusively focusing on the faults and foibles of how media workers use these systems, the concept of responsible AI demands that AI-based systems and the companies that develop and deploy them (see \cite{Kak2023Make}) adhere to societal values and human rights and prevent detrimental outcomes \cite{Dignum2019Responsible}. Particularly for journalists and news organizations, where AI brings forth both unique challenges and opportunities \cite{Fridman2023How}, there is a pressing demand for further exploration into the creation of ethical media technology guardrails. As the Council of Europe argues in the publication of a recent set of guidelines for newsrooms, “news organizations should have procedures in place to recognize, and where feasible, assess and mitigate risks that result from the way journalistic AI systems are implemented” (p. 13). This approach aims to optimize the advantages of AI for news organizations and the public at large (e.g., automation of repetitive tasks, assistance for in-depth reporting) while preventing adverse impacts (e.g., bias in reporting and undermining public trust in journalism) \cite{Aissani2023Artificial} \cite{Helberger2022}. As \cite{Shin2022Countering} (p. 900) further put it, “regulating for responsible AI also means regulating for the responsible use, implementation, oversight, control, and contestation of AI systems.” 

However, the effective contestation and oversight of AI systems in journalism demand a nuanced understanding of both the operational dynamics within news organizations and the areas where journalists may lack the required knowledge to challenge or mitigate the influence of ‘unruly AI.’ Yet, it is essential to acknowledge that most newsrooms are likely to adopt off-the-shelf AI products from dominant US-based platform companies, and they will have limited capacity to modify or influence these technologies directly \cite{Dodds2023Popularity-driven}, highlighting a power imbalance between newsrooms and platform companies \cite{Simon2022Uneasy}. This reality underscores the necessity of not just governing journalists’ interactions with AI but also reforming the broader institutional practices within news organizations with respect to AI. Without such reforms, efforts to foster responsible AI use risk being reduced to mere window dressing, failing to address the more profound systemic challenges AI systems bring forth in journalism. 

In response to these challenges, various news organizations have initiated steps to address the ethical implications of AI \cite{Becker2023Policies}. Wired, for instance, was one of the first to establish ground rules for using generative AI, emphasizing ethical usage and transparency. The magazine decided against publishing stories with text generated or edited by AI, except when AI’s involvement is central to the story \cite{Tobitt2023ethics}. As \cite{Cools2023Towards} put it, some of these guidelines may “help map, measure, and manage risks, offering ways to efficiently develop policies for mitigating risks that are identified.” This proactive approach reflects a broader awareness among media organizations about the ethical ramifications of AI in journalism. 

Against this backdrop of technological awareness, financial limitations, and a lack of information-sharing among different teams inside newsrooms, our research aims to explore the following question: How do knowledge silos affect the adoption of responsible AI practices in journalism? We argue that this question is crucial because understanding current applications and journalistic information-sharing practices about AI systems is essential to developing effective strategies and policies that ensure the beneficial use of AI in journalism without compromising ethical standards and journalistic integrity. Drawing from 14 interviews with journalists in the Netherlands, this paper examines current AI applications and debates inside four major Dutch newsrooms. We specifically chose these given the Netherlands’ reputation as an early adopter of AI systems and having an overall progressive stance on digital innovations \cite{Vergeer2020Artificial}. While our findings from the Dutch context may not be universally generalizable, this case study offers valuable insights, especially in terms of ethical AI usage and the potential for scaling responsible practices within the global journalistic community. 

\section{Adopting Responsible AI Practices in Journalism}
\subsection{Defining AI and AI Use in News Organizations}
Artificial intelligence is difficult to define, but in practice, it typically takes the form of ‘narrow’ computer systems that focus on specific tasks and problems usually associated with human abilities \cite{Broussard2019Artificial} \cite{Mitchell2019Artificial}. AI systems employ a variety of techniques – often from the machine learning branch in computer science and statistics – varying in complexity, autonomy, and abstraction. Within the news industry, AI “serves as an umbrella term to encompass a range of technologies and is commonly understood by many as the computational simulation of human activities and skills in specific domains” \cite{Simon2023AIa}. As such, artificial intelligence is increasingly shaping journalistic work. A growing number of journalists and news workers nowadays use AI to automate tasks (such as translation or transcription of content) or aid in other news tasks, for example, investigative work \cite{Fridman2023How}. News organizations also employ AI on the distribution side, for example, to optimize paywalls and recommend content to users \cite{Hansen2023AI-} \cite{Simon2023AIb}.

However, the increasing adoption of AI in the news industry is not a smooth process and has been going hand in hand with a range of concerns. Many journalists have expressed worries about AI potentially taking over their jobs, also referred to by some as the ‘robocalypse’. Additionally, the adoption of AI has brought existing tensions and power dynamics in news organizations to the fore \cite{Simon2024Artificial}. As the societal and journalistic impact of AI is growing, so do concerns around the safety, fairness, responsibility, power, and ethics of artificial intelligence systems and applications \cite{Dörr2017Ethical}, for example, observe a significant shift in the locus of responsibility within news production, extending beyond the individual journalist to various stakeholders, including algorithms, service providers, and data collectors. Others have noted tensions between the use of AI and journalistic ideals, including public service, autonomy, and objectivity \cite{Milosavljević2021‘Our}, concerns regarding the growing dependency on technology companies and the effects on (editorial) autonomy due to AI \cite{Dodds2023Popularity-driven} \cite{Simon2022Uneasy}, and tensions between economic interests and commitments to journalistic integrity and quality. \cite{Wiley2023Grey}, for example, has examined this from a legal perspective, noting that algorithmic journalism often has to navigate ‘fuzzy’ situations, with legal concerns arising in all parts of news production and distribution, especially around the use of data, defamation liability, and individual privacy. 

\subsection{Responsible AI}

Efforts to address and mitigate these challenges regarding the use of AI are often grouped under the label ‘Responsible AI.’ While it lacks a clear definition, the term is often used to describe specific organizational choices and practices around AI and refer to the emerging area of AI governance, which focuses on the ethical and accountable development, deployment, and use of artificial intelligence systems. It emphasizes the need for AI technologies to align with societal and professional values, respect human rights, and avoid harmful consequences, and is pursued in parallel by governments, academia, industry, civil society, and, increasingly, news organizations. Many of these are working on establishing best practices and regulatory frameworks to promote or enforce the responsible development and deployment of AI technologies, as well as considering the broader implications and potential risks associated with AI applications and taking the steps necessary to mitigate them. The unifying trend among many of these initiatives is a focus on accountability, responsibility, and transparency (ART) as essential facets of ethical AI systems \cite{Krausova2022Disappearing}, with further emphasis on the ability to explain outcomes, human responsibility in design and deployment, and transparency in technical solutions, data, and involved stakeholders.

While the concept of ‘Responsible AI’ is fairly novel in journalism studies, there is a thematic alignment with the concept in parts of the literature, with scholars discussing, among other things, issues such as algorithmic bias, data privacy, transparency, impact on employment, and the potential for AI-generated content to be indistinguishable from human-written content, in the context of algorithmic journalism and AI. Three distinct categories that repeatedly feature in the wider literature on digitalization and automation in the news focus on (1) upholding professional values inherent to journalistic practices, such as objectivity and accuracy, (2) adhering to legal frameworks governing technological applications (and AI) in journalism and (3) considering democratic obligations associated with journalism and the societal impact of AI in news. 

Previously mainly a niche endeavor, Responsible AI has grown in importance following the proliferation of generative AI and the resulting widescale bottom-up adoption of AI. With user-friendly interfaces like those of OpenAI’s ChatGPT or Google’s Gemini, powerful AI technology has become accessible to every journalist. Although emerging ethical guidelines have started to outline the principles of responsible AI use, they often remain (too) abstract and lack specific implementation strategies. This also highlights the importance of understanding how journalists in the field perceive and practice responsible AI.

\section{Information Sharing and Knowledge Silos}

Journalism’s primary purpose is to impart accurate and relevant information and “to provide people with the information they need to be free and self-governing” \cite{Shapiro2010Evaluating}. In this context, journalists have long come together as Communities of Practice (CoPs) to learn from and collaborate with one another \cite{Wenger1998Communities} both intra- and extra-organizationally. CoPs are based on mutual engagement and unite actors of different abilities, sharing “a repertoire of communal resources (routines, sensibilities, artifacts, vocabulary, styles, etc.) that members have developed over time” (\cite{Wenger1998Communities}, p. 98). Sharing knowledge within journalistic CoPs ideally ensures that reporters stay well-informed and up-to-date and can produce comprehensive and diverse news stories, thus contributing to maintaining high journalistic standards and a well-informed society. 

Specializations naturally develop in organizations as a response to increasing complexity, enabling different actors to focus on specific tasks and areas of expertise. For instance, network scientists and journalists have distinct roles that require different skill sets and knowledge bases. However, the problem arises when there is no infrastructure to facilitate the exchange of critical information between these specialized units. This lack of communication and coordination leads to knowledge silos, where specialized teams operate in isolation, unaware of or unable to integrate the knowledge and insights from other units. 
Journalists rarely operate independently. Instead, they are embedded in formal and informal structures \cite{Clement2017Searching}, with the social context of a news organization (Singer, 2004) shaping their work. Research into knowledge management has demonstrated that, among other things, such organizational settings can both enable and constrain knowledge sharing – not just between journalists but also between the newsrooms and other parts of the organization. Such ‘knowledge silos’ – “individuals or higher-level collectives that serve as heterogeneously distributed repositories of knowledge” (\cite{Phelps2012Knowledge}, p. 1117) arise in many organizations, in some cases on purpose. Especially in large organizations, operating in silos and compartmentalization (for example, departments) can lead to greater efficiency or make functioning possible in the first place \cite{Waal2019Silo-Busting:}.

Yet, despite the lack of empirical studies on organizational and knowledge silos \cite{Bento2020Organizational}, the general view, including in journalism studies, is that knowledge silos are often an obstacle to innovation. For example,  a ‘silo mentality’ – the presence of barriers to communication and exchange \cite{Bento2020Organizational} – in legacy structures or organizational designs can discourage collaboration \cite{Waal2019Silo-Busting:}. Effectively, knowledge silos can manifest themselves in four different forms within organizations: As (1) individual silos, where certain individuals possess valuable knowledge or expertise but do not share it, (2) as departmental silos, where different departments or teams within an organization hoard information and fail to communicate effectively with each other, (3) as technological silos where information is stored in different software systems, databases, or file formats that are not easily accessible or integrated across the organization, making it more difficult for employees to access or share relevant information across different systems and limiting the interoperability of the newsroom, and (4) cultural silos where organizational culture (in addition to organizational structure) discourages collaboration, information sharing and learning. 
Furthermore, knowledge silos may also manifest in various structural dimensions within (news) organizations, including vertical, horizontal, internal, and external forms. Vertical silos refer to communication barriers that emerge along the organizational hierarchy, impeding the flow of information and interaction between upper management and lower-tier employees. Conversely, horizontal silos describe the knowledge disconnects that occur among peers who, despite operating at the same hierarchical level, possess divergent specialties, thus hampering interdisciplinary understanding and cooperation (see \cite{Finnegan2006Knowledge}). At the same time, internal silos are recognized within the confines of a singular organization, where they inhibit collaboration and the sharing of information across different divisions or teams. External silos, in contrast, arise between organizations and their overarching parent entity, restricting the transfer of knowledge and strategic congruence between the two (see \cite{Hullova2019Independent}).

The effects can be problematic. Knowledge silos are often blamed for hindering innovation, decision-making, and overall organizational performance. They are also seen as impeding the flow of information, limiting the cross-pollination of ideas within the organization, and preventing the organization from capitalizing on the collective intelligence of its staff. Within a journalism context, this has been shown to matter, for example, in the context of how editorial decision-making is automated. In this context, differing levels of knowledge and power can be unequally distributed within organizations \cite{Drunen2022Safeguarding}, and are often limited to particular teams in newsrooms \cite{Milosavljević2021‘Our}. In newsrooms, such team compartmentalization may lead to a lack of cross-team collaboration, reduced efficiency in information sharing, and slower response times to emerging stories. For instance, if the editorial team responsible for producing news articles is unaware of insights gathered by the investigative team working on the same topic, this can result in incomplete or fragmented reporting, potentially affecting the depth and accuracy of the coverage. Additionally, in digital media organizations, if the social media team is not aligned with the content creators, there might be missed opportunities for timely and effective dissemination of news, thereby reducing the organization’s overall impact and reach.

To overcome knowledge silos, it is generally suggested that organizations should foster a culture of collaboration, implement effective knowledge management systems, encourage information sharing, and promote cross-functional communication and learning, with some studies indicating that there is a positive relationship between the application of such a “silo-busting” approach and the extent of learning and knowledge exchange, the quality and outcomes of collaboration and the overall strength of the organization \cite{Waal2019Silo-Busting:}.

\section{Methods}
We theorize that knowledge silos and the lack of information sharing also affects the adoption and implementation of responsible AI practices. Consequently, this study aims to explore if, and if so, how different dimensions of knowledge silos (i.e., individual, departmental, technological, and cultural) affect the adoption of responsible AI practices in journalism. We focus on four major media outlets in the Netherlands to answer this question.

The Netherlands forms an excellent case study for examining the adoption of AI in newsrooms due to its proactive stance and substantial investment in AI across various sectors, including media and journalism. The country’s government and private sector have collaborated actively through initiatives like the Netherlands AI Coalition, highlighting a national commitment to responsible and innovative AI development \cite{Vergeer2020Artificial}. The Dutch approach to AI, which focuses on ethical, legal, and societal aspects (ELSA), underpins a framework conducive to studying AI’s impact on journalism and newsmaking. Furthermore, the Dutch media landscape’s participation in global AI initiatives and adherence to stringent AI regulations provides a unique context for us to explore AI’s role in journalism. While other countries with similar AI advancements could also be interesting to study, we selected the Netherlands because its specific regulatory and ethical framework offers valuable insights into the dynamics of responsible AI adoption in journalism.	

The four news organizations chosen for this study are \textit{de Telegraaf}, \textit{de Volkskrant}, the \textit{Nederlandse Omroep Stichting} (NOS), and \textit{RTL Nederland}. These organizations were selected due to their significant presence in the Dutch media landscape and their existing engagement with AI technologies in journalism. They also represent a broad spectrum of the Dutch media landscape, encompassing print, digital, and broadcast media, and are pivotal in understanding the role of knowledge silos in AI adoption in newsrooms. For instance, commercial broadcaster \textit{RTL Nederland}, alongside \textit{NPO}, the Dutch public broadcaster, has publicly pledged to adhere to ethical guidelines when using AI systems for news production (Priestley, 2021). Mediahuis, the Belgian media company that owns \textit{de Telegraaf}, has also started experimenting with AI tools like ChatGPT for tasks such as writing marketing messages and job descriptions, illustrating a pragmatic approach to AI utilization inside news media \cite{Majid2023How}.

We conducted 14 semi-structured interviews with editors, managers, and journalists across the abovementioned newsrooms. The respondents were approached via LinkedIn, and their selection was based on the criterion that they use AI in their daily practice. Relying on snowball sampling (Parker et al., 2019), agreeable participants were then asked to recommend others who fit the research criteria and who potentially might be willing to participate, and so on.
We conducted an equal number of interviews inside \textit{de Telegraaf} ($n$ = 4), \textit{de Volkskrant} ($n$ = 4), and \textit{NOS} (n = 4). Due to availability, we conducted fewer interviews inside \textit{RTL} ($n$ = 2). Interviews were conducted using a semi-structured interview instrument developed based on the research question and the existing literature (see appendix). We chose this approach for its flexibility and depth, allowing for a comprehensive exploration of the participants’ experiences and attitudes toward AI in journalism. The semi-structured format enabled the interviewees to express their views freely while ensuring that all relevant topics were covered \cite{Gubrium2001} \cite{Magaldi2020Semi-structured}.

As strict privacy was a condition for access to each of the newsrooms, we decided to anonymize all our participants. In the results, we refer to “editors” without making a distinction between e.g., “Editor-in-Chief” or “Online Editor.” We provide more information about the participants in the text when the quotes cannot be used to identify a particular individual. Research ethics approval was granted by [ANONYMISED]. 

The interviews were conducted in Dutch and translated into English for coding. Drawing on the principles outlined by \cite{SkjottLinneberg2019Coding}, this study involved a three-tiered coding process, starting with open coding to identify emerging concepts and themes from the data. This was followed by focused coding, which aimed at detecting patterns and trends within these initial findings. Finally, axial coding was employed to establish consistency and associations between the identified categories. This approach facilitated a comprehensive understanding of the complex phenomena being studied, ensuring a thorough and nuanced qualitative data analysis. 

Coding was carried out by one team member, after which the initial coding was checked by a second team member. Where necessary, labels were refined and adjusted. Regular communication, code reviews, and adherence to a shared coding style guide were standard practice during the study. This collaborative coding effort contributed towards the consistency and agreement among team members in interpreting and implementing coding standards, ensuring high inter-coder reliability.

\section{Results}
Our results reveal that knowledge silos regarding the responsible use of AI in newsrooms can manifest themselves in several forms. We identified four different types of knowledge silos (vertical, horizontal, external, and internal) playing a role across three different areas of the newsrooms related to the way news workers use and understand AI: professional limitations, infrastructural limitations, and ethical guardrails. We show how these different types of silos manifest in various areas of the newsroom and condition the way journalists approach Responsible AI practices in journalism, but also how they impact collaborative relationships with colleagues across their news organizations. 

\subsection{Knowledge Silos and Professional Limitations}

The challenges of knowledge silos within professional environments are multidimensional, as highlighted by our interviews with newsroom personnel. During our interviews, editors often mentioned a knowledge gap between themselves and colleagues with different technical backgrounds. We have named these “horizontal silos” and “vertical silos.” These silos impede the seamless integration of technology and collaboration across different levels and areas of experience within an organization. 

The notion of horizontal silos is exemplified by an editor from \textit{de Telegraaf}. While discussing the integration of machine learning and natural language processing tools into their daily tasks, the \textit{de Telegraaf} editor stated: “Oh, I find it very difficult to explain. [My colleague, an IT technician] probably knows more about that. He knows about numbers.” Interestingly, while this editor suggests a knowledge barrier between the editorial and technical staff, this admission also highlights a disconnect, not only in understanding but in confidence, when it comes to the domain of machine learning, which is already in use in the editor’s newsroom. It is telling that the editor associated technological expertise with proficiency in numbers, suggesting a stereotypical view that may further entrench these silos. We found something similar when interviewing an editor from \textit{RTL}, who claimed, “I am sitting here, in my management office, thinking about what the editors should do about AI, while my journalists, of course, already have a much better understanding of these tools from using them.” These findings point to a possible need to pay attention to how horizontal silos could be impacting the processes of production inside newsrooms in a way that makes the development of new technologies opaque to certain sections of the newsroom.

This example could also suggest the need to bridge the gap between technical and non-technical staff to foster a more collaborative and informed workplace environment. This is important because we also find that terminology could become a barrier for internal vertical silos. For example, when we asked an editor from \textit{RTL} to name some of the AI technologies that they are currently using, he responded:

\begin{quote}
    I don’t know if what I have in my head right now falls under either of those [AI technologies]. I’m just not good at the terminology yet… I will name one [AI technology], and then you have to tell me whether it fits.
\end{quote}

As this quote shows, editors might be aware of new technologies being adopted in newsrooms. Yet, they typically leave determining whether these technologies qualify as AI to colleagues they perceive to have greater technical expertise.
 
We found that knowledge gaps also exist between different departments within the same company. For example, a respondent from \textit{de Volkskrant} noted a clear demarcation between departments handling private data and the editorial team, indicating a lack of transparency and understanding about data privacy practices and AI tools implementation across departments: 

\begin{quote}
    When it comes to handling that private data, yes, I never use users’ private data. That’s good, right? There is a separation between the advertising and sales department and the editors. I don’t think we actually do [collect users’ private data] as editors ourselves. So, if and when that happens, it happens in another department.
\end{quote}

This quote is particularly interesting because, from an external viewpoint, the organization bears collective responsibility for adherence to legal frameworks, such as the General Data Protection Regulation (GDPR), that regulate the use of users’ data. However, based on these interviews, internally, there is a prevalent attitude that assigns responsibility to other departments of individuals within the organizations, encapsulated in the mindset of ‘it is not our concern, but theirs.’ This delineation of accountability, or rather a displacement thereof, signifies an issue of responsibility fragmentation within news organizations, impeding not only the flow of knowledge but also the coherent application of regulatory compliance.
 
The professional backgrounds required to deal with AI innovations in the newsrooms are sometimes so different that they often require professionals from different departments to come together. For example, \textit{de Volkskrant}’s attempt to develop LOOKER, an in-house web data tracker, showcases how different teams, like News Analytics and Directional Analytics, need to come together to collaborate. As one of the journalists involved in the project put it:

\begin{quote}
    LOOKER. There’s a whole team behind it. That is really a huge number of people. There is a News Analytics Team, which really focuses on the newsroom and news analysis. There's a Directional Analytics Team that focuses on strategic numbers and things like that. They made that themselves, but it is based on existing data systems, such as Google Analytics in the end.
\end{quote}

Furthermore, we also found knowledge silos between levels of positions and, thus, a vertical one. For example, editors, in general, are less aware of the actual tools that are being used in the newsrooms but are more aware of the trends of AI adoption. Conversely, journalists are more up-to-date with on-the-ground AI tools. For example, a journalist who showed himself to be very tech-savvy during the interview and claimed to navigate the dark web to download illegal datasets that could help him with his reporting argued a big knowledge gap between him and his editors exists: “Well, to be honest, I think most of the editors have no idea what AI is, what it can do, and what we should be thinking about it.” 

Finally, the disconnect between parent and subsidiary companies, referred to as an external vertical silo in our study, further complicates the situation. An example from \textit{de Volkskrant} illustrates this gap, where the parent company \textit{DPG} is actively engaged in AI adoption, yet such initiatives are not as prevalent within the editorial office of the subsidiary. When a journalist from \textit{de Volkskrant} was asked about plans for AI adoption in the newsroom, she mentioned: 

\begin{quote}
    Well, not in the editorial office; I think at \textit{DPG}, undoubtedly it is. They are very busy working on that. But here at the editors’ offices – no. People think AI impacts a lot of what we do, but no.
\end{quote}

Up to this point, our results seem to indicate that knowledge silos impact the responsible implementation of AI in the newsroom by creating obstacles to AI literacy and learning opportunities. Interestingly, some interviewees suggested that cross-departmental knowledge-sharing practices can break knowledge silos. When asked about their knowledge of AI, a journalist at \textit{NOS} admitted that although they “wouldn’t know exactly how the technology behind AI tools work,” the more technical editors could “explain it very well.” However, we observe how a culture of sharing tools, techniques, and skills is emerging within the tech journalism community, promoting a more innovative and collaborative approach to journalism. 

These results leave the door open to argue that addressing these various forms of knowledge silos require concerted efforts to foster open communication, continuous education, and cross-functional teams. By doing so, news organizations can create an environment where technology is not divisive but a unifying force that enhances the journalistic process and products. 

\subsection{Breaking Silos}

The integration and effective use of AI in the newsroom is an evolving challenge that requires a well-developed organizational infrastructure. Through our interviews with industry professionals, we identified three critical elements that significantly impact the adoption and perception of responsible AI in journalism: in-house training, tool testing, and team collaboration. These components are not just operational necessities but are foundational to fostering an environment where AI can be used ethically and effectively to enhance journalistic practices \cite{Council2023}.

Firstly, in-house training emerges as an important factor in preparing newsrooms for the challenges and opportunities presented by AI. Our interviews revealed that such training programs are instrumental in building a more informed team of journalists. For instance, a journalist from \textit{de Volkskrant} shared that their parent company, \textit{DPG}, offers courses on AI applications in journalism: 

\begin{quote}
    Within \textit{DPG}, there is a kind of teaching program for journalists. And that’s where people who know more about navigating the web, for example, come and teach us. Recently, there was also something from the VVOJ [Vereniging van Onderzoeksjournalisten, Association of Investigative Journalism] and something specifically about AI and how you can use it in music research. We recently made a production in which we trained an algorithm ourselves.
\end{quote}

This education initiative empowers journalists, providing them with a greater sense of agency and competence in leveraging AI for innovative projects. Secondly, tool testing appears to be another important aspect that newsrooms must consider. It involves ensuring AI-based applications meet journalistic standards of responsibility and integrity. An editor from \textit{de Telegraaf} emphasized that the adoption of automated tools is not merely a technical decision but a journalistic one, reflecting on the ethical implications of using AI to condense articles: 

\begin{quote}
    The question of automated journalism is not a question of fairness; it is a question of journalistic responsibility. For example, if we have a web article that is 700 words long, but we only have space for 500 words for the paper… if there is a tool that makes it shorter automatically if there is a good tool for this, we might experiment with it. However, that test should focus on whether it is acceptable and responsible from a journalistic point of view. And if that tool is so good that it is journalistically acceptable and justifiable, then we can make use of it. And is that unfair to journalists? No, that’s not unfair. It helps them. Then, the reporter, who has been somewhere all morning to make a story, does not have to cut back that story at eight o’clock. So, it also makes the life of the journalist a bit more pleasant.
\end{quote}

These initiatives not only allow journalists to make informed choices about the digital tools they employ, ensuring these technologies enhance their work without compromising journalistic values, but they also necessitate a substantial level of AI literacy. Where such expertise is not inherent, it encourages collaboration between editorial staff and technical teams, thereby helping to break down knowledge silos and foster a multidisciplinary approach to tech adoption in journalism. 

Thirdly, we found that the composition of teams and room for experimentation within news organizations plays a crucial role in the swift and effective adoption of emerging technologies such as AI. An editor from \textit{RTL Nieuws} noted the rapid pace of development of tools like ChatGPT, stressing the need for dedicated time and resources to explore these advancements thoroughly: “[ChatGPT] is a development that is going so fast that we need to give ourselves time to dig into this, so we don’t get caught off guard with how the market is already handling it.” Similarly, \textit{de Telegraaf}’s approach to forming specialized teams for evaluating AI tools exemplifies how structured collaboration between editorial and product departments can lead to more strategic technology adoption. According to one of the editors at \textit{de Telegraaf}: 

\begin{quote}
    These conversations take place between the editorial team and our product department, which possesses the most insightful understanding of the tools available to us. Through these discussions, we organize demonstrations with various providers to assess which tools best suit our needs and decide how to implement them effectively. This process ensures that we have thoroughly deliberated on what actions to take and what to avoid, thereby making informed decisions.
\end{quote}

These teams not only assess the suitability of various AI applications but also ensure that their integration into the newsrooms is thoughtful and informed. Clearly, the successful integration of AI in journalism is not solely dependent on the technology itself but also on the infrastructural support provided by the organization. In-house training, tool testing, and team composition are not just administrative tasks; they are essential practices that enable journalists to navigate the complexities of AI and avoid the generation of knowledge silos inside news organizations. 

\subsection{Guidelines and legal frameworks}

he implementation of responsible AI in newsrooms requires the development of journalistic guidelines and legal frameworks to ensure its responsible use. Despite this need, our research indicates that the use of and responsibility for AI remains compartmentalized, creating silos within organizations. Regulatory frameworks aimed at privacy and transparency, such as the GDPR, are infrequently mentioned during the interviews, highlighting a gap in integrating these critical measures into daily journalistic practice. 

We did, however, encounter positive exceptions during our interviews. Consider the experience of an editor at \textit{de Volkskrant} who expressed unease regarding the collection and use of personal data and the transparency of the AI tools used in her newsroom, particularly for advertising:

\begin{quote}
    Transparency is key; we never utilize private data, and that's a positive aspect, wouldn't you agree? We also maintain a clear distinction between our advertising and sales departments and the editorial team. It's true that ads can be intrusive, and they reflect what users have interacted with. Unfortunately, advertising is integral to our business model—it's essential for generating the necessary revenue. We're always looking for opportunities to include ads, but we must balance this with user experience. If ad intensity becomes overwhelming and drives our users away, it defeats the purpose.
\end{quote}

Then, the interviewee continued:

\begin{quote}
    At \textit{DPG Media}, we are committed to full compliance with all relevant legislation. I've encountered concerns from readers in the past who were reluctant to accept cookies, voicing their complaints to us. My consistent response has been that any data we store is handled strictly in accordance with the law. \textit{DPG Media} upholds these standards meticulously. Should you have any further technical or legal inquiries, I recommend reaching out to our legal department for detailed assistance. As a journalist, I'm acutely aware of how bothersome these issues can be, and I empathize with your frustration.
\end{quote}

When faced with contestable data collection methods, legal compliance is often the default resolution to disputes, as one interviewee suspected the AI tools were “not so transparent” but trusted in an unnamed “European legislation” to ensure legitimacy. The problem is compounded by the absence of a solid conversation around internal guidelines for AI tool usage, at least in the newsrooms where our research occurred. A tech journalist at \textit{de Volkskrant} admitted to a lack of discussion around guidelines, “there’s not a very solid conversation or any kind of guidelines,” underscoring a significant issue: The difficulty journalists face in grasping the emerging challenges posed by AI. 
\
These findings emphasize the need for a concerted strategy to address the knowledge silos within news organizations, fostering an environment where AI can be adopted responsibly and transparently. 

\section{Discussion and conclusion}
Adopting AI responsibly inside news organizations is not easy, as it requires a concerted effort to bridge different perspectives, interests, and the voices of technicians, editors, journalists, and managers \cite{Helberger2022Towards}. This collaboration between the various actors involved in news production is fundamental to harnessing the benefits of artificial intelligence while preventing the potential risks associated with these systems \cite{Beckett2023Generating}. However, among the many challenges that could impact information sharing around the responsible use of AI inside news organizations are knowledge silos, where information is isolated within one part of the organization and not easily shared with others \cite{Waal2019Silo-Busting:} \cite{Drunen2022Safeguarding}.

Our analysis identified ‘knowledge silos’ as a significant barrier to the effective integration of AI within news organizations. The term ‘silos’ refers to the segmentation and isolation of information, expertise, and communication within different departments or groups, leading to inefficiencies and obstacles in collaborative efforts. This concept, rooted in organizational behavior and knowledge management theories \cite{Finnegan2006Knowledge} \cite{Phelps2012Knowledge}, highlights the challenges posed by fragmented structures that inhibit the flow of information and shared learning.

This study aimed to explore if, and if so, how, knowledge silos affected the adoption of responsible AI practices in journalism through a cross-case study of four major Dutch media outlets. Our findings underscore the existence of such silos across the Dutch media outlets we examined, coupled with a lack of organizational and more structural conditions to enable knowledge sharing, which presents a significant barrier to the effective operationalization of responsible AI initiatives. 

Our results show that knowledge silos regarding the responsible use of AI in newsrooms can take several forms. In this article, we identified four different types of knowledge silos (vertical, horizontal, external, and internal) playing a role across three distinct areas of the newsrooms related to the way news workers use and understand AI: professional limitations, infrastructural limitations, and ethical guardrails. 

In our study, we observed that these silos prevent cross-departmental collaboration and the holistic adoption of AI technologies, which require a unified approach to ethics, technology, and journalism practices. The presence of silos results in compartmentalized knowledge, limiting the organization's ability to leverage AI effectively and responsibly. By emphasizing this dynamic, our research sheds light on the need for integrative strategies that bridge these divides, fostering a more collaborative environment conducive to responsible AI innovation.

Differentiating silo types allowed us to pinpoint specific challenges. For example, and drawing from our interviews, we observed how vertical silos separate hierarchical levels, with editors lacking technical expertise often relegated to colleagues in the IT department or journalists covering technology. We also identified horizontal silos existing between departments, such as gaps between journalists working from different thematic sections inside the same newsrooms. 

Furthermore, we distinguished between internal silos that exist within a news organization and external silos that impede knowledge sharing with external stakeholders, such as parent companies or regulatory bodies. This distinction was beneficial to recognize the challenges that news organizations will have to face in the future because while internal silos could be addressed through knowledge-sharing programs, to solve external silos, news organizations will have to rely on industry-wide projects to bring different stakeholders together.

By analyzing how these different types of silos manifest in various areas of the newsroom—professional limitations, infrastructural limitations, and ethical guardrails—we gain a deeper understanding of how journalists approach Responsible AI practices in journalism. For instance, a horizontal silo between journalists and data scientists could lead to journalists feeling apprehensive about using AI tools due to a lack of understanding. This is a particularly important conclusion because our results showed that although there are initiatives aiming to solve this situation (e.g., education, tool testing), there is also a cultural undercurrent where those lacking knowledge hold beliefs that they are not able to learn. Conversely, a vertical silo between editors and IT staff could result in editors feeling they lack the authority to enforce responsible AI practices. Additionally, knowledge silos can also impact collaborative relationships with colleagues across news organizations \cite{Wenger1998Communities}. For example, an internal silo separating the legal department from the editorial team could lead to confusion and delays when implementing ethical frameworks for AI use. Unknown legal and social risks are thus the inevitable downside of AI innovations in the context of silos. 

According to our interviews, it is important to notice that the implication of knowledge silos extended beyond questions of mere inefficiency. As our results suggest, these silos can create a breeding ground for accountability evasion. Aligned with prior studies \cite{Dörr2017Ethical}, our interviews revealed that journalists often defer responsibility for the ethical and responsible implementation of AI in the newsroom to the tech or the legal team. This case hinders the need to move from the traditional independent role of journalists to a more collaborative role that includes communication and knowledge transfer among different stakeholders to ensure the responsible use of AI. Furthermore, a lack of shared understanding across departments could exacerbate existing biases within AI algorithms, as those with limited technical knowledge may be unable to challenge potential biases within the systems effectively. Our results reinforce the notion that responsible AI requires a “many hands” approach \cite{Helberger2018Exposure} to dismantle silos and encourage cooperation and knowledge exchange. This study, however, also points to the need for media organizations to think more structurally about how to organize and facilitate knowledge sharing. This should also include creating room and incentives for knowledge sharing.

Our research aligns with existing literature highlighting the importance of “bridging the gap between ideals and practices” \cite{Schiff2020Principles} regarding responsible AI inside newsrooms. Journalistic principles are only as effective as their practical application \cite{Dodds2022Newsroom}, and knowledge silos can significantly impede this translation. For instance, based on our interviews, robust ethical frameworks for AI use in newsrooms become meaningless if siloed departments lack the knowledge or capacity to implement them. 

By fostering collaboration and knowledge exchange across departments, news organizations can create a more fertile ground for developing responsible AI practices. This requires not only training programs but also a fundamental shift in organizational culture, creating incentives and dedicated spaces for knowledge sharing. 

The limitations of this study also offer valuable insights for future research. Focusing on four Dutch news outlets limits the generalizability of our findings. Further research exploring a more comprehensive range of media organizations across different geographical regions is necessary to determine if these siloed structures and challenges are universal or context-specific. Additionally, while this study focused on knowledge silos, other organizational structures, and cultural dynamics may also hinder responsible AI practices. Investigating these factors alongside silo effects would provide a more comprehensive picture of the challenges faced by newsrooms. 

\newpage
\bibliography{sn-bibliography}

\end{document}